\documentstyle{mn}
\input{epsf}
\newif\ifAMStwofonts
\def\GPT{{\cal G}_{\rm PT}}
\def\Gd{{\cal G}_{\delta}}
\def\PPT{\phi_{\rm PT}}
\def\ii{{\rm i}}
\def\d{{\rm d}}

\def\xib{\overline{\xi}}

\def\ltsima{$\; \buildrel < \over \sim \;$}
\def\simlt{\lower.5ex\hbox{\ltsima}}
\def\gtsima{$\; \buildrel > \over \sim \;$}
\def\simgt{\lower.5ex\hbox{\gtsima}}

\def\mG{{\cal G}}

\title[Scaling in Gravitational Clustering]
{Scaling in Gravitational Clustering, 2D and 3D Dynamics} 

\author[D. Munshi et al.]{D. Munshi$^1$, 
F. Bernardeau$^2$, A. L. Melott$^3$,
R. Schaeffer$^{2}$\\
$^1$ Queen Mary and Westfield College, London E1 4NS, United Kingdom \\
$^2$ CEA, Service de Physique Th\'eorique,
F-91191 Gif-sur-Yvette c\'edex, France\\
$^3$Department of Physics and Astronomy, University of Kansas,
Lawrence, Kansas 66045, U.S.A.}

\pagerange{000,000}
\pubyear{1996}

\begin{document}

\maketitle

\begin{abstract}
Perturbation Theory (PT) applied to a cosmological density field 
with Gaussian initial fluctuations suggests a specific hierarchy for 
the correlation functions when the variance is small.
In particular quantitative predictions have been made for the
moments and the shape of the one-point probability distribution function
(PDF) of  the top-hat smoothed density.
In this paper we perform a series of systematic checks of these
predictions against N-body computations both in 2D and 3D
with a wide range of featureless power spectra.
In agreement with previous studies, we found that the reconstructed
PDF-s work remarkably well down to very low probabilities,
even when the variance approaches unity. 
Our results for 2D reproduce the features for the 3D dynamics. In 
particular we found that the PT predictions are more accurate 
for spectra with less power on small scales. 

The nonlinear regime has been explored with various tools, PDF-s,
moments and Void Probability Function (VPF). These studies
have been done with unprecedented dynamical range, especially 
for the 2D case, allowing in particular more robust determinations
of the asymptotic behavior of the VPF. We have also introduced 
a new method to determine the moments based on the factorial 
moments. Results using this method and taking
into account the finite volume effects are presented.

\end{abstract}


\section{Introduction}

The general frame of the current theory for the large--scale
structure formation is generally believed to be based on the 
gravitational amplification of small density perturbations.
The detailed study of such a mechanism started in the seventies 
with a complete treatment of the linear growth of the initial 
fluctuations (e.g. Peebles 1980), but progress towards an understanding
of the nonlinear aspects of the dynamics has been very slow.
Zel'dovich (1970) proposed an interesting approximation for a 
qualitative description of some nonlinear aspects but this
approximation cannot provide accurate quantitative predictions, and
is certainly not valid in the late stage of the nonlinear dynamics.
The discovery of the self-similar solutions with the assumption
of stable clustering (Davis \& Peebles 1977) provided
a valuable insight into the fully nonlinear regime.
Subsequent numerical analysis, however, partially confirmed the
existence of such a regime but only for very large mean density
fluctuations, for rms density fluctuations above ten. 
More recently a lot of theoretical and numerical efforts
have been devoted to the study of the quasi-linear regime corresponding to
density fluctuations below unity.
These results suggest we should  distinguish three different regimes:

\begin{enumerate}
\item
{the linear or quasi-linear regime at large scale when the variance 
is below unity;}
\item
{the intermediate regime when the variance is between unity and 10;}
\item
{the nonlinear regime for the smallest scales for which the rms density
fluctuation exceeds 10.}
\end{enumerate}

The nonlinear regime is expected to be reasonably well
described by the self-similar solutions, which is very useful because
it provides a well-defined frame for theoretical predictions.
We recall here the general results. For an initial power-law
spectrum, $P(k)\propto k^{n}$, the small scale two-point correlation
function, $\xi_2$, is expected to follow a power-law behavior,
\begin{equation}
\xi_2(r)\sim r^{-\gamma}
\end{equation}
with an index $\gamma$ related to  $n$ (Davis \& Peebles 1977).
Moreover the higher order correlation functions are also expected
to follow  a specific behavior for their global scale dependence,
\begin{equation}
\xi_p(\lambda\ r_1,\dots,\lambda\ r_p)=
\lambda^{-\gamma(p-1)}\xi_p(r_1,\dots,r_p).
\end{equation}
From this general property it is possible to show that the average
$p$-point correlation functions,
\begin{equation}
\xib_p={1\over V^p}
\int_V\d^3r_1\dots\int_V\d^3r_p\ \xi_p(r_1,\dots,r_p),
\end{equation}
in a cell of volume $V$ are related to the average two-point
correlation function with
\begin{equation}
\xib_p=S_p\ \xib_2^{p-1}\label{scal},
\end{equation}
where the $S_p$ coefficients are scale independent. Note that $\xib_p$
is identical to the $p$-order cumulant of the one-point density 
probability distribution function. There are however no definitive theories
for the $S_p$ parameters, although different
models have been proposed (Hamilton 1988,
Fry 1984, Schaeffer 1984, Balian \& Schaeffer 1988, 1989a,
Bernardeau \& Schaeffer 1992, Munshi \& Padmanabhan 1996).

The intermediate 
regime is certainly the most poorly understood from a theoretical
point of view. It has been the subject of only few semi-analytic
investigations (Jain, Mo \& White 1995, Mo \& White 1996, Padmanabhan 1996, 
Munshi \& Padmanabhan 1996).

The linear and quasi-linear regime have been investigated in details 
over the last few years using perturbation theory techniques. In particular 
it is possible to show that for Gaussian initial conditions  the 
average $p$-point correlation functions follow the hierarchy (Fry 1984, 
Goroff 	et al. 1986, Bernardeau 1992),
\begin{equation}
\xib_p=S_p^{\rm PT}\ \xib_2^{p-1},\label{scal2}
\end{equation}
where the $S_p^{\rm PT}$ 
coefficients (not necessarily identical to the $S_p$ 
in eq. [\ref{scal}]) depend on the local shape of the power
spectrum (Goroff et al. 1984, Bouchet et al. 1992). 
A lot of effort has been devoted recently to the analytic calculation 
of these coefficients in different cases
(Juszkiewicz, Bouchet \& Colombi 1993, Juszkiewicz et al. 1995,
Bernardeau 1994a). Thus $S_3$ and $S_4$ are known 
analytically or semi-analytically
for a Gaussian filter and a power-law spectrum (\L okas et al. 1995).
More interesting is the case of a top-hat filter for which the whole 
series of the coefficients $S_p$ is known for any cosmological
model and any power spectrum in 3D (Bernardeau 1994b) and
in 2D (Bernardeau 1995).
The fact that the complete series is known allows us to build the shape
of the one-point density Probability Distribution Function (PDF).
These predictions have been checked in peculiar cases at various
levels, for the moments (Bouchet et al. 1992, \L okas et al. 1995, 
Baugh, Gazta\~naga \& Efstathiou 1995, Colombi et al. 1995)
or for the shape of the PDF (Bernardeau 1994b). All these checks 
have been made for the 3D dynamics.

Applications to observational data have, so far, given contradictory results.
For instance Gazta\~naga \& Frieman (1994),
Gazta\~naga (1995) concluded that the observed 
galaxy distribution in the APM survey
reproduces the theoretical predictions, but Bernardeau (1995)
claimed a significant discrepancy. In any case the fact that
quantitative predictions, that are direct consequences
of the gravitational instability scenario, could be checked 
in observational data boosted the theoretical investigations in this
domain. In particular the calculation of the next-to-leading order
term in the perturbative expansion of $\xib_p$ have been
contemplated by Scoccimarro \& Frieman (1995a, b), \L okas et al. (1995).
In this paper we propose a systematic checks of the PT results
for power law spectra, completing previous results in 3D and
investigating the 2D case. Our objective is to circumvent
the validity range of all the PT results at the level of the PDF-s,
and also for the void probability function.

The paper is divided as follows. In section 2 we recall the mathematical 
tools that originally have been developed for the fully nonlinear regime
by Balian \& Schaeffer (1989a, b), and turn out to be
useful for the quasi-linear regime as well.
In the subsequent section we present the specific results obtained in
the quasi-linear regime, and compare them to the numerical results
in quasi-linear and nonlinear regime. Most of the mathematical details
about calculating $S_p$ parameters using method based on factorial 
moments are presented in the appendix.

\section{Scaling and Counts in cells Statistics}

To explore the statistical properties of the particle distribution
in an N-body simulation we consider the Count Probability
Distribution Function (CPDF), $P_l(N)$,
the probability of having $N$ particles in a spherical cell of
radius $l$. In order to relate the CPDF to the $p$-point correlation
functions
we can consider its generating function, 
\begin{equation}
P(\lambda)=\sum_{N=0}^{\infty}\,\lambda^N\ P_l(N),
\end{equation}
that can be shown to be given by (White 1979, Schaeffer 1984, 
Balian \& Schaeffer 1989a, Szapudi \& Szalay 1993),
\begin{equation}
P(\lambda)=\exp\left[
\sum_{p=1}^{\infty}{(n\,V)^p (\lambda-1)^p\over p!}
\ \xib_p\right],
\end{equation}
where $n$ is the mean number density of particles in the considered sample.
Assuming scale invariant $p$-point correlation
functions we can further write,
\begin{equation}
P(\lambda)=
\exp\left[-{\phi((1-\lambda)\,n\,V\,\xib_2)\over \xib_2}\right],
\end{equation}
where the function $\phi$ is defined by,
\begin{equation}
\phi(y)=-\sum_{p=1}^{\infty}S_p\ {(-y)^p\over p!},
\end{equation}
(we have set $S_1=S_2=1$). We can notice that the Void Probability 
Function (VPF) is equal to the generating function for
$\lambda=0$. As a result the probability $P_l(N)$
can be derived from the VPF through the relation,
\begin{equation}
P_l(N)\equiv\left.
{\partial^N\,P_l(N)\over\partial\lambda^N}\right\vert_{\lambda=0}
={(-n)^N\over N!}\ {\partial^N\ P_l(0)\over \partial n^N}\label{Pdev}
\end{equation}
where the partial derivatives are taken at fixed volume.
From eq. (\ref{Pdev}) one can show that
for a continuous field 
\begin{equation}
p(\delta)\,\d\delta=\d\delta\,
\int_{-\ii\infty}^{+\ii\infty}
{\d y\over 2\pi\xib_2}
\,\exp[-\phi(y)/\xib_2+(1+\delta)\,y/\xib_2].\label{Linv}
\end{equation}
For small $\xib$ {\it and} very small $\delta$,
$p(\delta)$ reduces to the well-known
Gaussian form, $p(\delta)\propto\exp(-\delta^2/2\xib)$, but
even for moderate small values
of $\delta$, strong deviations from the Gaussian behavior are expected
(Balian \& Schaeffer 1989a, Bernardeau 1992).

In the nonlinear regime, under the assumption of scale-invariance
of the correlation functions (in the sense that the coefficients $S_p$
are scale-independent), $P(N)$ 
and $p(\delta)$ exhibit characteristic scaling laws. 
This implies that the VPF has a specific scale dependence, namely
that, 
\begin{equation}
\sigma(N_c)=-\ln[P_l(0)]/(n\,V),\end{equation}
is a function of the combination
\begin{equation}
N_c=n\,V\,\xib_2\end{equation}
only. The only reasonable physical models are those for which
$\sigma(N_c)$ decreases to 0 when $N_c$ is large (Balian \& Schaeffer 1989a).
It is thus quite reasonable to assume that, at large $N_c$,
$\sigma(N_c)$ follows a power-law behavior,
\begin{equation}
\sigma(N_c)\sim a\,N_c^{-\omega},\end{equation}
where $\omega$ is a model dependent parameter that should
be comprised between 0 and 1.

Let us describe in more details the consequences of these
assumptions in the fully nonlinear regime as they were found by Balian
\& Schaeffer (1989a). Thus, for the shape
of $P_l(N)$ three domains are expected. They are delimited by
$N=N_c$ and $N=N_v$, with
\begin{equation}
N_v=N_c\,(\xib_2/a)^{-1/(1-\omega)}.
\end{equation}
This number is smaller than $N_c$ when the variance is
large (we will see that $a$ is of the order of 1).
Then, when $N<N_v$, $P_l(N)$ is expected to follow a specific scaling,
\begin{equation}
P_l(N)={1\over N_v}\,g\left({N\over N_v}\right)\label{Pg}
\end{equation}
where the function $g$ is given by
\begin{equation}
g(z)=-z^{-1/\omega}\,\int_{-\ii\infty}^{+\ii\infty}
{\d t\over2\pi\ii}\,\exp\left[z^{1(1-\omega)/\omega}\,
\left(t-t^{1-\omega}\right)\right].
\end{equation}
It depends on $\omega$ only, and when $z$ is small 
it reads,
\begin{eqnarray}
g(z)&\approx& {1\over z}\,
\sqrt{{1-\omega\over2\pi\omega}\,
\left[{1-\omega\over z}\right]^{(1-\omega)/\omega}}\times\nonumber\\
&&\,\exp\left(-\omega\left[{1-\omega\over z}\right]^{(1-\omega)/\omega}\right).
\end{eqnarray}
When $N_v<N<N_c$, $P_l(N)$ is expected to follow a power-law
behavior,
\begin{equation}
P(N)={a\over N_c\,\xib_2}\,{1-\omega\over\Gamma(\omega)}\,
\left({N\over N_c}\right)^{\omega-2}.
\end{equation}
And when $N>N_c$ another scaling behavior is expected,
\begin{equation}
P_l(N)={1\over N_c\,\xib}
h\left({N\over N_c}\right),\label{Ph}
\end{equation}
with
\begin{equation}
h(x)=-\int_{-{\rm i}\infty}^{+{\rm i}\infty}
{\d y\over 2\pi\,\ii}\,\phi(y)\,\exp(x\,y).
\end{equation}
Furthermore, it is quite natural to expect that $\phi(y)$
has a singular behavior for a negative and small value of $y$,
\begin{equation}
\varphi(y)\sim\varphi_s-a_s\,\Gamma(\omega_s)\,(y-y_s)^{-\omega_s}
\end{equation}
This singularity induces an exponential cut-off for $h(x)$,
\begin{equation}
h(x)\sim a_s\,x^{\omega_s-2}\exp(-\vert y_s\vert\,x),
\end{equation}
when $x$ is large. the scaling functions $\sigma(N_c)$,
$g(z)$ and $h(x)$ have been studied in great detail
computationally including spurious effects that may alter their
measurements in finite $N$-body catalogues (Bouchet \& Hernquist 1992,
Colombi et al. 1992, 1994, 1995).

\section{Scaling Parameters from the Quasi-linear Regime}

In the quasi-linear regime the scaling (\ref{scal2}) 
for the dominant part of the correlation functions is a consequence 
of the dynamical evolution. The picture obtained in the previous section
for the shape of $P_l(N)$ is however
no more valid when $\xib_2$ is small. But actually
the functions $h(x)$ and $g(z)$ can still be used
to describe the shape of the density PDF-s. The main difference with
the highly non-linear regime is that the 
low and large density cut-offs merge together, thus suppressing
the domain of the power law behavior. It is then still relevant 
to describe the results in terms of $a$, $\omega$ for the low
density domain and in terms of $a_s$, $y_s$ for the large density
tail.

We take advantage here 
of the fact that, for top-hat filtering, the whole series of the $S_p^{\rm PT}$
parameters is known (Bernardeau 1994b) through their generating function
$\phi_{\rm PT}(y)$.
Note that $\PPT(y)$ can be seen as a low $\xib_2$ limit of a more general
function $\phi(y,\xib_2)$. More precisely it means that
whatever $y$,
\begin{equation}
\phi(y,\xib_2)\to\PPT(y)\ \ {\rm when}\ \ \xib_2\to 0,
\end{equation}
but there is no guarantee that this convergence is uniform, that
is that the limiting function $\PPT(y)$ is reached at the same
time in terms of low values of $\xib_2$ whatever $y$.
It is however strongly suggested by the numerical results
(Bernardeau 1994b, Colombi et al. 1996), and we will make
this assumption in the following, thus considering the
global properties of $\PPT(y)$ as a valid model for the
calculation of the scaling functions, $\sigma(N_c)$, $h(x)$, $g(z)$.

So the function, 
$\PPT(y)=-\sum_{p=1}^{\infty}\ S_p^{\rm PT}\ (-y)^p/p!$,
is given by the system,
\begin{eqnarray}
\PPT(y)&=&y+y\,\GPT[\tau(y)]+{1\over 2}\tau^2(y)\nonumber\\
\tau(y)&=&-y\,{\d\GPT[\tau(y)]\over \d\tau}\label{imp}
\end{eqnarray}
where the function $\GPT(\tau)$ can be deduced from the spherical
model dynamics, since we have 
\begin{equation}
\GPT(\tau)=\Gd\left[\tau\,{\sigma_M\left(M_0[1+\GPT(\tau)]\right)\over
\sigma_M(M_0)}\right]
\end{equation}
where $\Gd(\tau)$ gives the quasi-linear density contrast as a function
of the linear density contrast, $\tau$, and $\sigma_M$ is the
rms density fluctuation at a given {\em mass} scale, $M_0$
being the mass scale associated with the filtering radius.
These results are valid for the 2D and 3D dynamics.
For power-law spectra,
\begin{equation}
P(k)\sim k^{n},
\end{equation}
the generating function $\GPT(\tau)$ reads,
\begin{equation}
\GPT(\tau)=\Gd\left(\tau\,\left[1+\GPT(\tau)\right]^{-(n+d)/(2 d)}\right).
\label{smoothG}
\end{equation}
To get quantitative predictions from PT one needs to know the expression
of $\Gd(\tau)$ in both cases.
There are no simple analytical expressions for $\Gd(\tau)$. It is 
however possible to develop $\Gd(\tau)$ with respect to $\tau$
and one obtains,
\begin{equation}
\Gd(\tau)=-\tau+{17\over21}\,\tau^2-{341\over567}\,\tau^3+
\dots
\end{equation}
for the 3D case (Bernardeau 1994b) and
\begin{equation}
\Gd(\tau)=-\tau+{18\over21}\,\tau^2-{29\over42}\,\tau^3+
\dots
\end{equation}
for the 2D case (Bernardeau 1995). 
One can then compute the values of the first $S_p$ coefficients
for the 3D case,
\begin{eqnarray}
S_3^{3D}&=&{34\over7}-(n+3),\\
S_4^{3D}&=&{6712\over1323}-{62\over3}(n+3)+{7\over3}\,(n+3)^2,
\end{eqnarray}
and for the 2D case,
\begin{eqnarray}
S_3^{2D}&=&{36\over7}-{3\over2}(n+2),\\
S_4^{2D}&=&{2540\over49}-{33}(n+2)+{21\over4}\,(n+2)^2.
\end{eqnarray}
To have more global properties of $\phi(y)$ it is necessary to know the
global shape of $\Gd(\tau)$ and one can  actually show that,
\begin{equation}
\Gd(\tau)=\left(1-{\tau\over\nu}\right)^{-\nu}-1,\label{apG}
\end{equation}
with
\begin{equation}
\nu={3\over2}\ \ \ {\rm for}\ \ \ 3D,\end{equation}
and
\begin{equation}
\nu={\sqrt{13}-1\over2}\approx1.3\ \ \ {\rm for}\ \ \ 2D,
\end{equation}
is actually a good approximate function. The parameters $\nu$
are chosen to reproduce the exact asymptotic behavior of
$\Gd(\tau)+1$ for large $\tau$.
Note that the Zel'dovich approximation would have given $\nu=d$
in (\ref{apG}), (Bernardeau \& Kofman, 1995). It is interesting to remark
also that the asymptotic behavior of $\phi(y)$ for large
values of $y$, and thus the value of $\omega$, is 
given by the asymptotic behavior of $\Gd(\tau)+1$ for large $\tau$
(Bernardeau \& Schaeffer 1992). From (\ref{smoothG}) it is easy to show
that 
\begin{equation}
\GPT(\tau)+1\sim\tau^{-{\nu/(1-\nu(n+d)/(2 d))}}
\end{equation}
so that 
\begin{equation}
\omega^{\rm PT}={d\over d(2+\nu)/\nu-(n+d)}.
\end{equation}
As a result for the 3D case we have,
\begin{equation}
\omega^{\rm 3D}={3\over 7-(n+3)},
\end{equation}
and for the 2D case,
\begin{eqnarray}
\omega^{\rm 2D}&=&{2\over 2(3+\sqrt{13})/(\sqrt{13}-1)-(n+2)}\\
&\approx&{2\over 5.1-(n+2)}.\nonumber
\end{eqnarray}
Note that these results are exact, in the sense that they do not depend
on the approximation (\ref{apG}) made for $\Gd(\tau)$.  These parameters
entirely determine the shape of the function $g(z)$.
It can be however more interesting to consider a more general
expression given by the saddle point approximation in eq. (\ref{Linv}),
\begin{eqnarray}
p(\delta)\d\delta&=&-{\d\delta\over\GPT'(\tau)}
\left[1-\tau\,\GPT''(\tau)/\GPT'(\tau)\over2\pi\sigma^2\right]^{1/2}\,
\times\nonumber\\
&&\exp\left(-{\tau^2\over2\sigma^2}\right),
\ \ {\rm with}\ \ \GPT(\tau)=\delta.\label{spoint}
\end{eqnarray}
This expression reduces to (\ref{Pg}) when the functions $\GPT'(\tau)$
and $\GPT''(\tau)$ are replaced by their asymptotic power-law behavior.

To obtain the position of the cut-off in $h(x)$ one needs to know the 
dominant singular
value of $\PPT(y)$ and the behavior of $\PPT(y)$ near this value. It is
actually not given by the singularity appearing in the expression
of $\Gd(\tau)$ but generically by the 2$^{\rm nd}$ equation of the system 
(\ref{imp}).
It is quite easy to see that there is a value of $y$ for which
$(\d y/\d\tau)=0$. At this point we have
\begin{equation}
\PPT(y)=\phi_s+r_s\,(y-y_s)+a_s\,(y-y_s)^{3/2}+\dots\label{sing}
\end{equation}
As a result we get an asymptotic shape for $h(x)$ given by\footnote{In
Bernardeau (1992), the regular term, $r_s\,(y-y_s)$, was neglected
so there is a slight change in eq. (\ref{hc})},
\begin{eqnarray}
h(x)&\approx&{3\,a_s\,\sigma\over 4\sqrt{\pi}}\,
(\delta+1-r_s)^{-5/2}\times\,\nonumber\\
&&\exp[-\vert y_s\vert(1+\delta)/\sigma^2+
\vert\phi_s\vert\,/\sigma^2].\label{hc}
\end{eqnarray}
The parameters $\phi_s$, $a_s$, $y_s$ and $r_s$ are given in table 1
for the 3D case as a function of the power law index $n$ and
in table 2 for the 2D case.

\begin{table}
\caption{Parameters of the singularity, eq. (\ref{sing}), for the 2D PT case}
\label{tab2}
\begin{tabular}{@{}lcccc}
$n$&$y_s$&$\phi_s$&$r_s$&$a_s$\\
-2& -0.171979& -0.19731& 1.60005& -1.71747\\
-1.5& -0.211979& -0.251915& 1.80602& -2.24518\\
-1& -0.276897& -0.349585& 2.22626& -3.40954\\
-0.5& -0.402865& -0.580901& 3.54813& -7.73243
\end{tabular}
\end{table}

\begin{table}
\caption{Parameters of the singularity, eq. (\ref{sing}), for the 3D PT case}
\label{tab1}
\begin{tabular}{@{}lcccc}
$n$&$y_s$&$\phi_s$&$r_s$&$a_s$\\
-3& -0.1848& -0.214286& 1.6565& -1.83824\\
-2.5& -0.213447& -0.253791& 1.80389& -2.20845\\
-2& -0.252692& -0.3113& 2.0344& -2.80188\\
-1.5& -0.309854& -0.402947& 2.44269& -3.93026\\
-1& -0.401244& -0.572949& 3.3437& -6.66931\\
-0.5& -0.57352& -1.00758& 6.63172& -18.6473\\
\end{tabular}
\end{table}

One can see that the singularity is sharper for low values
of $n$, and for the 2D case. Note that for $n\ge0$ there is no
singularity anymore, and the form (\ref{spoint}) 
only can be used to describe $P_l(N)$.
In such a case the asymptotic behavior of $\mG_{PT}(\tau)$
is
\begin{equation}
1+\mG_{PT}(\tau)\approx\left({\tau\over\tau_c}\right)^{2d/(n+d)},
\end{equation}
so that $\vert\tau_c\vert\approx 1.47$ for the 2D case and 
$\vert\tau_c\vert\approx 1.69$ for the 3D case.
As a result the large density tail takes the form,
\begin{eqnarray}
&p(\delta)\d\delta=\displaystyle{
-{\tau_c\,(n+d)\over 2d}\,(1+\delta)^{(d-n)/(2d)}}
\times\,\\
&\times \sqrt{2n/(n+d)}\,
\exp\left(-{\tau_c^2\over2\sigma^2}\,(1+\delta)^{(n+d)/d}\right)\ \d\delta
\nonumber
\end{eqnarray}
which gives a sharper cut-off than in the expression (\ref{hc}).

\begin{figure*}
\protect\centerline{
\epsfysize = 2. truein
\epsfbox[29 527 588 712]
{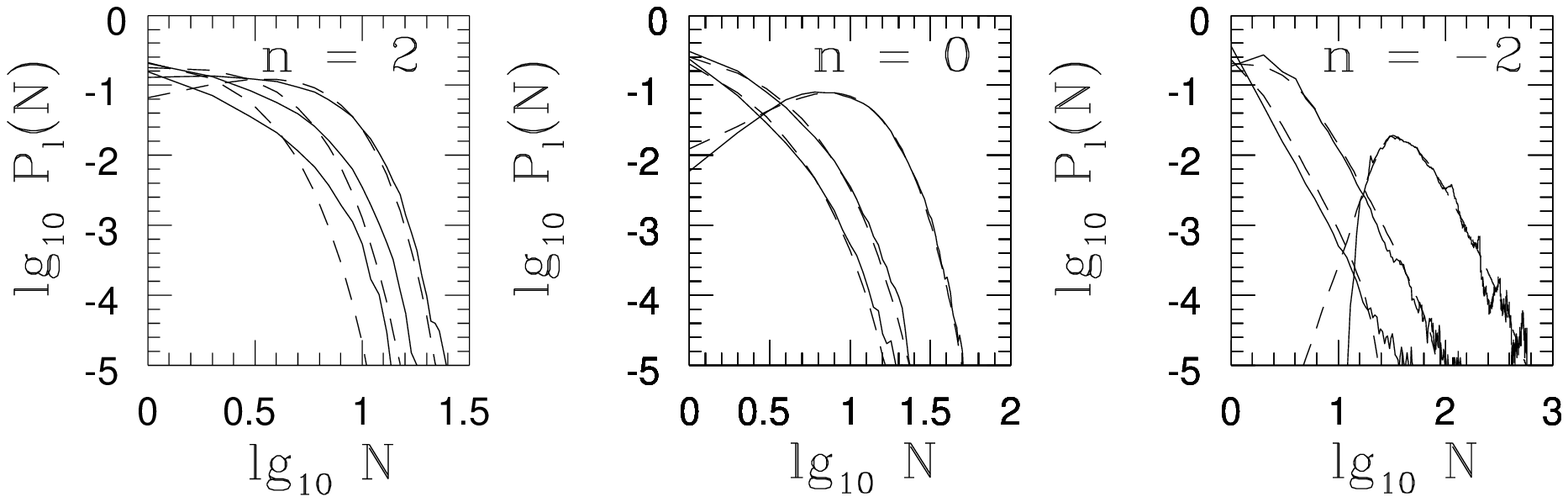}
}
\caption{The measured $P_l(N)$ in 2D N-body for simulations $n= 2$, $n=0$
and $n= -2$ compared to the theoretical prediction of PT
(long dashed lines)}
\end{figure*}

\begin{figure*}
\protect\centerline{
\epsfysize = 2.5 truein
\epsfbox[20 147 588 359]
{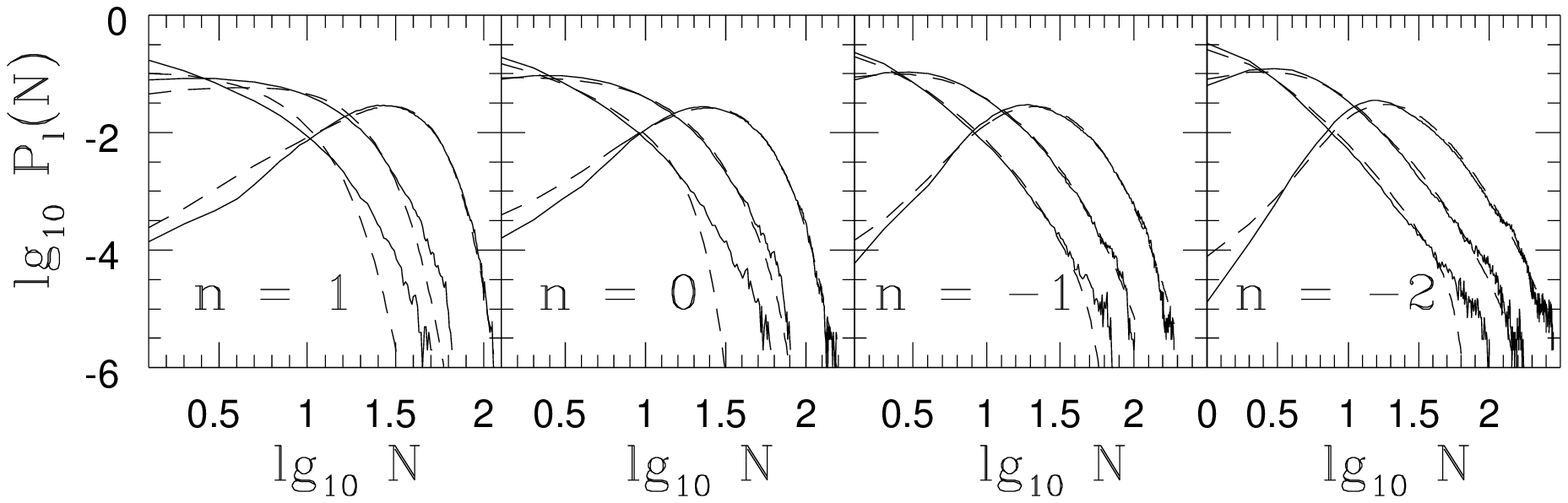}
}
\caption{The measured $P_l(N)$ in 3D N-body for simulations $n= 1$, $n=0$,
 $n= -1$ and $n= -2$ compared to the theoretical prediction of PT
(long dashed lines)}
\end{figure*}

\begin{figure*}
\protect\centerline{
\epsfysize = 2. truein
\epsfbox[24 512 588 705]
{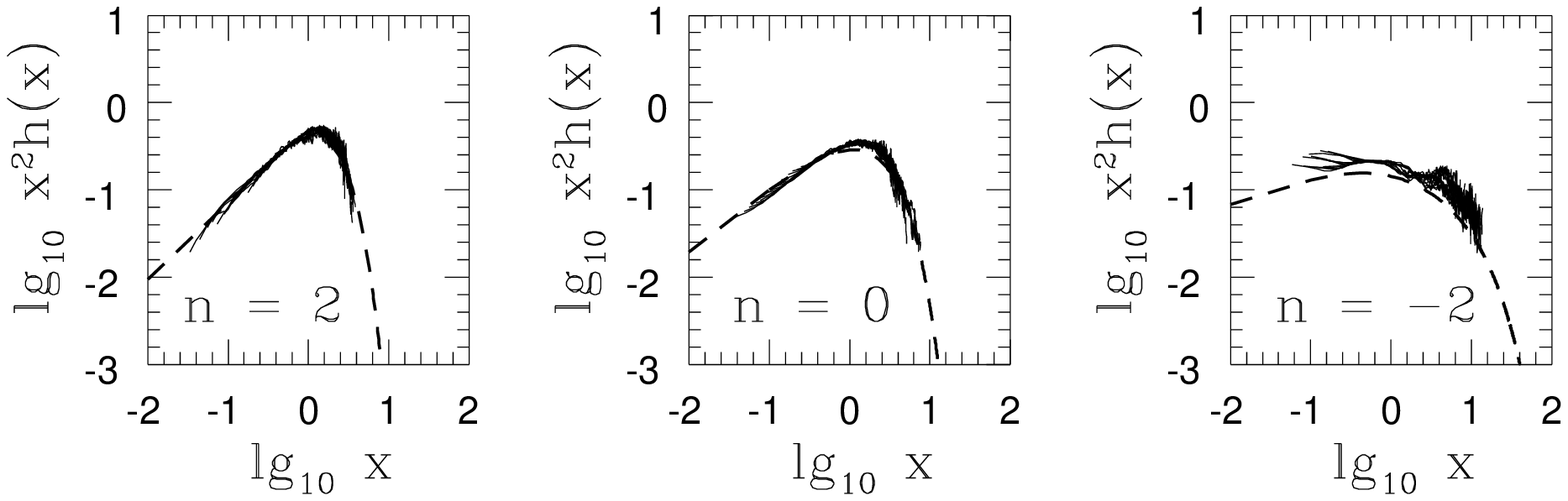}
}
\caption{The measured $h(x)$ in 2D N-body simulations for $n=2$, $n=0$
and $n=-2$ in highly nonlinear regime. Long dashed lines correspond to 
fitting function of the form (50) with the parameters of table 5.}
\end{figure*}

\begin{figure*}
\protect\centerline{
\epsfysize = 2.5 truein
\epsfbox[24 387 580 612]
{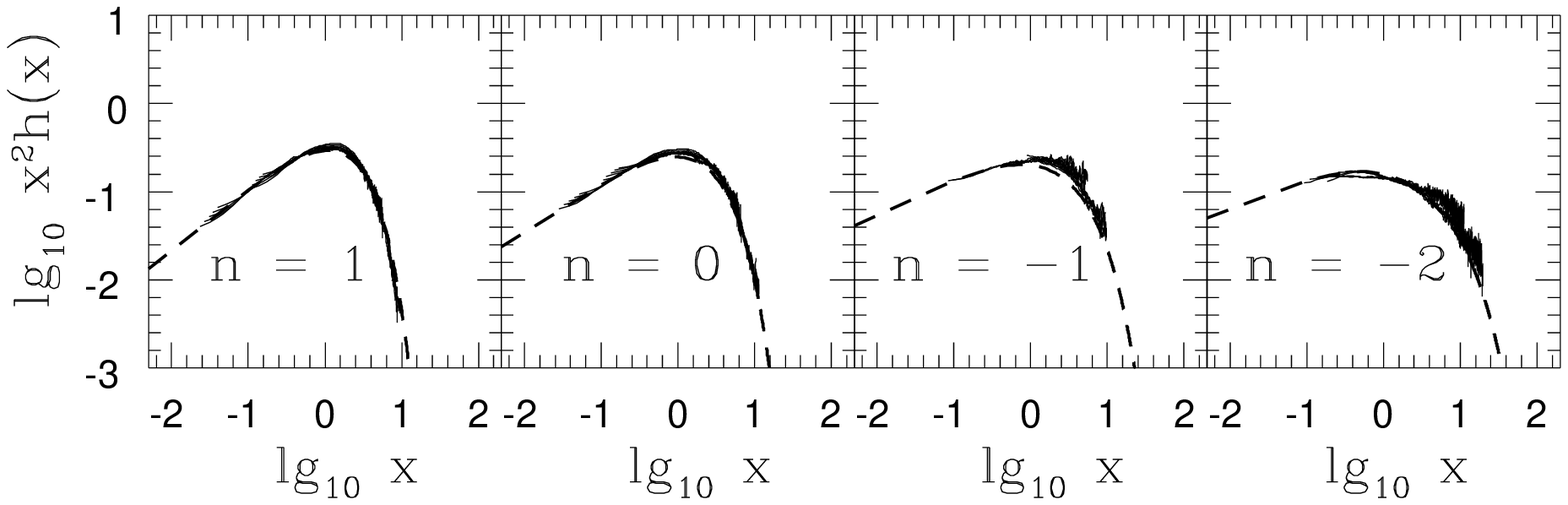}
}
\caption{The measured $h(x)$ in 3D N-body simulations for $n=1$, $n=0$
,$n=-1$ and $n = -2$ in highly nonlinear regime. 
Long dashed lines correspond to 
fitting function of the form (50) with the parameters of table 6.}
\end{figure*}

\section{Count probability distribution function
in Numerical Simulations}

\subsection{The simulations}

The simulations used here are numerical models for the gravitational
dynamics of  collisionless 
particles in an expanding background. We are studying evolution of initial 
Gaussian perturbations in $\Omega = 1$ universe. All the simulations are 
done with a particle-mesh (PM) code with $512^2$ particles in 2D in an 
equal number of grid points and $128^3$ particles with  $128^3$ grid points 
in 3D (Melott 1986; Melott, Weinberg \& Gott 1988, hereafter MWG). 
More details about
the peculiarities of the simulations used here can be seen in 
Melott \& Shandarin (1993). The code has at least twice the dynamical 
resolution of any other PM code with which it has been compared.

We use a subset of initial conditions used by Beacom et al. (1991) 
and Kofman et al. (1992) in their studies.  The time evolution of 
the N-body models can be seen in the video accompanying the
paper of Kauffmann \& Melott (1992). The models for which we carry out 
our analysis are featureless power-law spectra of the general form,
\begin{eqnarray}
P(k)&\propto&\ k^n\ {\rm for}\ k \le k_c,\\
&=&~ 0\ {\rm for}\  k > k_c.
\end{eqnarray}
We have analyzed power-law models with $n=$2, 0, -2 in 2D and 
$n=$1, 0, -1, -2 in 3D with a cutoff in each case at the Nyquist wave 
number: $k_c = 256\,k_f$ for 2D and $k_c = 64\,k_f$ for 3D where 
$k_f={2 \pi/ L_{\rm box}}$ is the fundamental mode associated with
the box size.

We choose $\sigma(k_{\rm NL})$, the epoch when the scale $2\pi/k_{\rm NL}$
is going nonlinear as a measure of time.

\begin{equation}
 \sigma ( k_{\rm NL}) = \left({{\int_{k_f}^{k_{\rm Ny}} P(k)\,k\,\d k } \over 
{\int_{k_f}^{k_{\rm NL}} P(k)\,k\,\d k }}\right)^{ 1 \over 2} 
\end{equation}

The first scale to go nonlinear is the one corresponding to the Nyquist 
wave number. This happens, by definition, when the variance
$\sigma$ is unity. 
Of course as $\sigma$
increases successive larger scales enter in the nonlinear regime.
The simulations were stopped at $\lambda_{\rm NL}=2l_{\rm grid},
\ 4l_{\rm grid},\ 8l_{\rm grid}, ....,\ L_{\rm box}/2$.
The growth rate of various modes in linear theory were studied in MWG for 
this PM code. The results given by our code 
at $ \lambda = 3l_{\rm grid} $ are equivalent 
to the ones obtain by 
a usual PM code at $\lambda = 8l_{\rm grid} $. This is due to
the staggered mesh scheme. So we expect that our code performs well at the
wavelength associated with four cells and since the collapse of 
$4\,l_{\rm grid}$-size
perturbations will give rise to condensations of diameter $ 2\,l_{\rm grid} $ 
or less, the smallest cell size
that we take into account should be bigger than $ 2\,l_{\rm grid} $.
On the other hand
Kauffman and Melott (1992) found that for voids of size greater than size 
$L_{\rm box}/4$ self-similarity was broken in a model equivalent to our 
index $n=-1$ 
in 3D, see also Gramann (1992) and Melott and Shandarin (1993).  
We therefore restrict our cell sizes 
to be less than $L/10$.  As a result our cell sizes vary between $2L_{grid}$ 
and $L_{\rm box}/10$.
We also do not use cells with $\sigma < 0.1$, since in this case shot noise 
becomes comparable to the fluctuation power impressed on the simulation.

Our simulations were started by using the Zel'dovich approximation 
(Klypin \& Shandarin 1983) but we wait
long enough before doing a comparison with the PT results so that
the effects of the Zel'dovich approximation have died away
(see for instance Baugh Gazta\~naga \& Efstathiou, 1995).

\subsection{The Count Probability Distribution Functions}

To evaluate the CPDF we calculate the
occupancy of spherical cells of size $l$ disposed on a regular mesh.
The number of cells to be used is to be as large as possible
so that all structures are fully taken into account. This can be achieved
by considering cells that are about $l_c$ apart where $l_c$
is the typical separation between particles in clusters 
($N_c(l_c) = 1$). Actually the major constraint comes from the resolution of
the $N$-body code. If $l_{\rm res}$ is the resolution scale of the N-body,
it is clear that probabilities below $(l_{\rm res}/L)^d$ are meaningless.
As a result we probe probabilities as low as $10^{-6}$ both in 2D and 3D.

Finite size of the sample affects mainly the large $N$ tail of the CPDF,
and this effect is all the more important that the cell size $l$
is large. The main effect is that the large density tail of the PDF
is dominated by a single cluster. It creates a bump in the PDF
which is followed by an abrupt cut-off of the distribution.
These features have been recognized as faked by Colombi et al (1995)
and methods to correct for it have been proposed. This is important in 
particular for the derivation of the high order moments that are
extremely sensitive to the defects in the large density tails.
In the method proposed by Colombi et al., this effect is corrected
using a theoretical prejudice, that is that the density PDF
is assumed to follow an exponential cut-off. This hypothesis is supported
by the PT theoretical results and by our understanding of
the nonlinear regime. We will use this method as well to compute the
high order moments.

We have followed the evolution of CPDF for all the spectra for different scales
with time. The different scales that we have studied are in the range 
$-2.2 \le \log(l/L_{\rm box}) \le -1.2 $ and they are separated by equal 
logarithmic intervals $ \Delta \log(l/L_{\rm box}) = 0.2 $. 

\subsubsection{Results in the Quasi-linear Regime}

In figs. 1 and 2 we present the results of the PDF-s for the 2D and 3D 
dynamics for different values of the variance. The latter is
at most of the order of $1.5$ (see table) and the
agreement is found to be extremely good for the smaller
variances. For variances approaching unity, the departure
from the PT results depends on the value of the initial index.

When there is a lot of power at large scale ($n$ small), the agreement
is better then for the other case. This is not too surprising
since when $n$ is large there is a lot of power at small scales
that can affect the behavior of the largest scales.

\begin{table}
\caption{Parameter values for the $P(N)$ in 2D dynamics}
\label{tabxi2D}
\begin{tabular}{@{}lccc}
                  &$n=2$&$n=0$&$n=-2$\\
\hline
$\bar\xi$&0.61&0.47&0.88\\
$\bar\xi$&1.04&1.00&1.32\\
$\bar\xi$&1.58&1.39&1.82\\
\hline
$n v$&4.11&10.33&69.55\\
$n v$&1.63&1.63&4.11\\
$n v$&0.65&0.65&0.65
\end{tabular}
\end{table}
\begin{table}

\caption{Parameter values for the $P(N)$ in 3D dynamics}
\label{tabxi3D}
\begin{tabular}{@{}lcccc}
                  &$n=1$&$n=0$&$n=-1$&$n=-2$\\
\hline
$\bar\xi$&0.40&0.48&0.57&0.71\\
$\bar\xi$&0.79&0.86&0.97&1.09\\
$\bar\xi$&1.44&1.48&1.5&1.64\\
\hline
$n v$&33.23&33.23&33.23&33.23\\
$n v$&8.33&8.33&8.33&8.33\\
$n v$&2.09&2.09&2.09&2.09
\end{tabular}
\end{table}

\subsubsection{Results in the nonlinear Regime}

The nonlinear regime has been explore mainly in terms of the 
function $h(x)$ (see eq. [\ref{Ph}]). Indeed when the variance is large
it is natural to expect that $N_c\,\xib_2\,P_l(N)$
is a function of $N/N_c$ only (for a given power law index). 
As a result one expects that, when plotted with the appropriate
variables, the PDF measured at for different smoothing
scales coincide. This test is presented on figs. 3 and 4.
Here we see that the locations of the PDF are indeed the same
when the adequate change of variable is made. It should be
noted however, that not all the PDF have been plotted. As the
function $h$ is pertinent is the large density tail only, the
PDF-s have been truncated in the law density domain.
We have removed part of $P(N)$ where it is dominated by shot noise.
 Typically $N<10$ while increasing with large scale power.
Very large $N$ part of $P(N)$ is dominated by large statistical
fluctuations, we applied smoothing to reduce such fluctuations.

In order to have quantitative results we used the parameterized 
fit proposed by Bouchet, Schaeffer \& Davis (1991) to describe
the function $h(x)$,
\begin{equation}
h(x) = {a ( 1 - \omega ) \over \Gamma( \omega)} {x^{\omega -2 }\exp( -x|y_s|)
\over ( 1 + b x )^c }\label{fith}
\end{equation}

The values of $\omega$ and $a$ are estimated from CPDF
(see in the next subsection), $\vert y_s\vert$ 
is found from
fitting the large $N$ tail of CPDF which shows an exponential cut-off.
The other two parameters are adjusted to reproduce the constraints $S_1 = 1$
and $S_2 = 1$ to at least $5$ percent accuracy.

\begin{table}
\caption{Parameters of the fitting function $h(x)$ (eq.\ref{fith})
for the 2D case}
\label{tabhx2D}
\begin{tabular}{@{}lccccc}
$n$&$\omega$&$a$&$\vert y_s\vert$&$b$&$c$\\
2&.85&3.35&1.31&3.2&-.83 \\
 0&.72&2.51&0.58&.06&1.2 \\
-2&.30& 1.2&0.09&1.1&.80 
\end{tabular}
\end{table}

\begin{table}
\caption{Parameters of the fitting function $h(x)$ (eq.\ref{fith})
for the 3D case}
\label{tabhx3D}
\begin{tabular}{@{}lccccc}
$n$&$\omega$&$a$&$\vert y_s\vert$&$b$&$c$\\
1& .70 & 2.38 & .64& .00& .00\\
0& .55& 1.59 & .41 & .38& .60\\
-1& .40 & 1.31 & .23& .65& .70\\
-2& .33 & 1.25 & .11& .80& .95 
\end{tabular}
\end{table}

We give the parameter values that have been used in tables
(\ref{tabhx2D}, \ref{tabhx3D}).

The variation with the power law index is extremely large,
in particular for the value of $\omega$. It is interesting to note
that the results are better when $n$ is large, for which 
there is a convincing overlapping of curves. The situation is
more questioning when $n$ is small, but it is still not 
clear whether it is due to some numerical difficulties
(finite volume effects are large when the power law index is small)
or to a genuine physical effect. In particular it has been 
found that the $S_p$ parameters reach their asymptotic value
for quite large values of the variance. The curve presented 
here may therefore still be in the intermediate regime for
which the function $h(x)$ is rapidly changing. 

We can also note that the CDM case as analyzed by 
Bouchet, Schaeffer \& Davis (1991) matches roughly with our calculations
of $n=-1$ or $n=-2$ 
for 3D dynamics with their computed value of $\omega = .4\pm .05$
and $\vert y_s\vert=0.08$.

\subsection{The Moments}


\begin{figure*}
\protect\centerline{ 
\epsfysize = 4.0 truein
\epsfbox[29 146 588 711]
{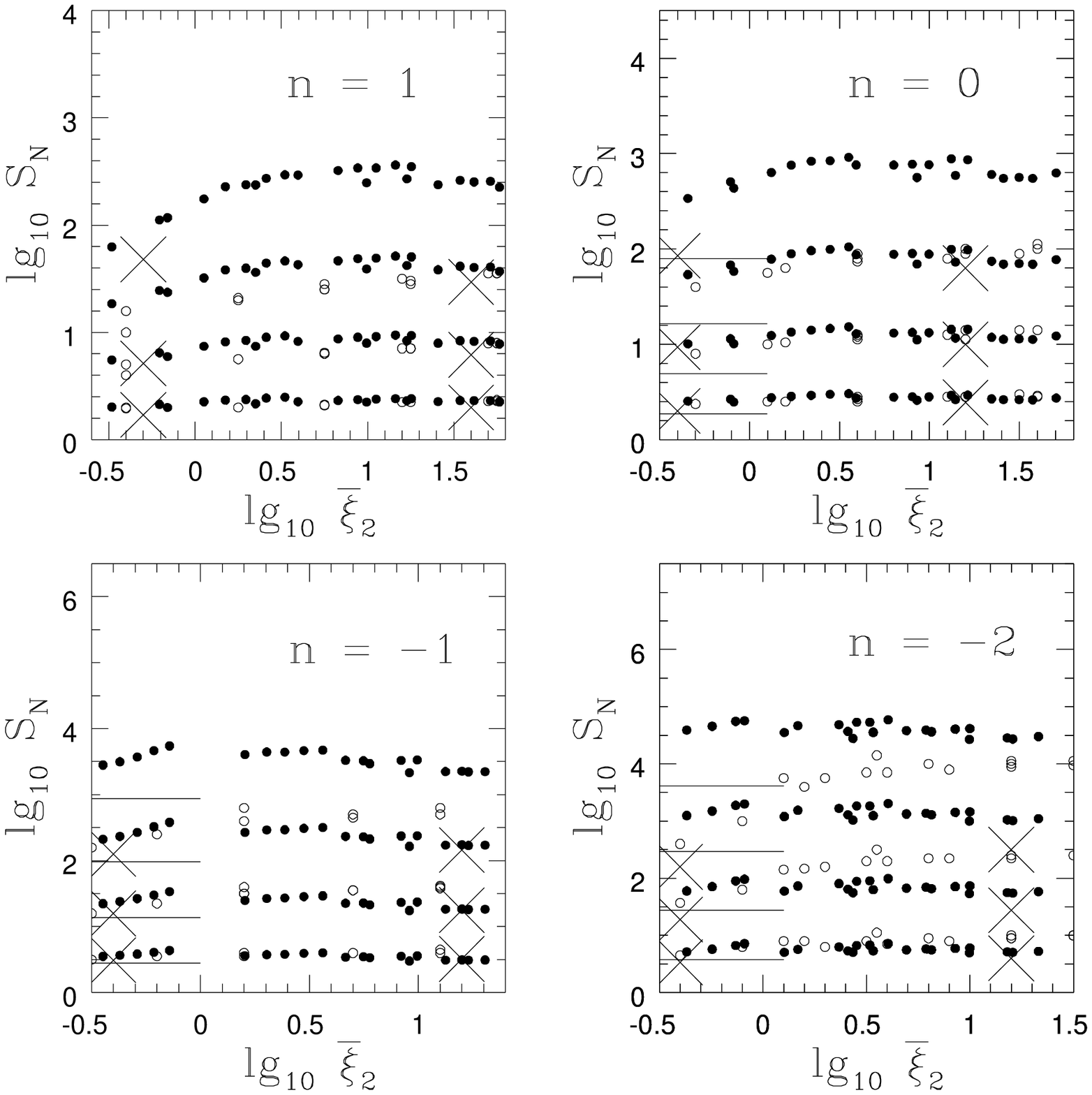}
}
\caption{The measured $S_p$ parameters in 3D N-body simulations 
for $n=1$, $n=0$, $n=-1$ and $n=-2$ are plotted against
$\bar{\xi}_2$. 
The filled dots our measurements 
of $S_p$ parameters for $p = 3$ to $p= 5$ after taking all the corrections.
Crosses represent  results from Lucchin et al. (1994). Open circles correspond
to measurements of $S_p$ parameters by Colombi et al. (1996). 
Solid lines  represent predictions from perturbation theory.  }
\end{figure*}

\begin{figure*}
\protect\centerline{ 
\epsfysize = 4.0 truein
\epsfbox[29 146 588 711]
{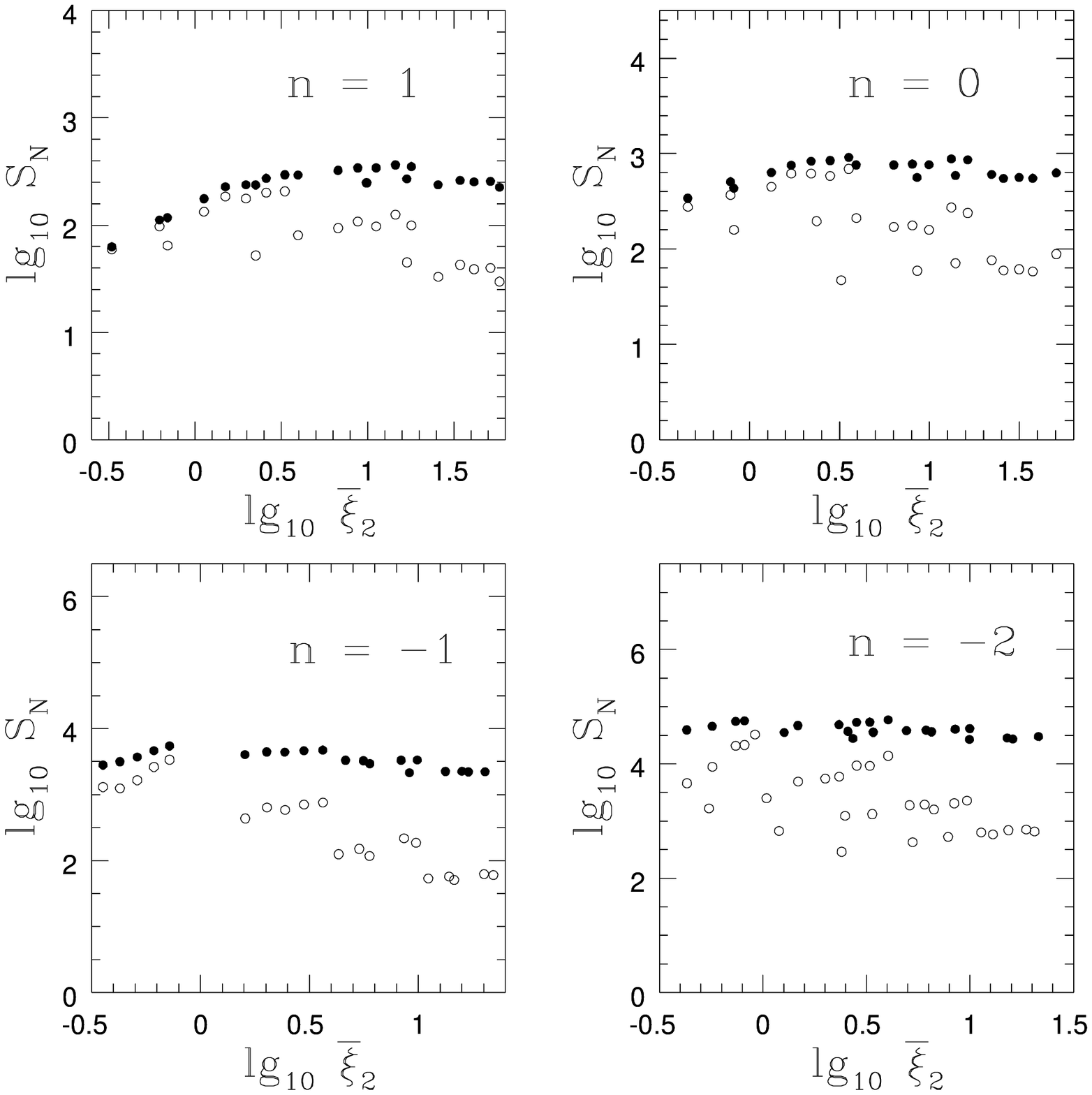}
}
\caption{The measured $S_6$ parameter after (filled circles) and 
before (open circles) finite volume corrections for different 
initial power spectra $n$. }
\end{figure*}


\begin{figure*}
\protect\centerline{
\epsfysize = 2.0 truein
\epsfbox[24 512 588 705]
{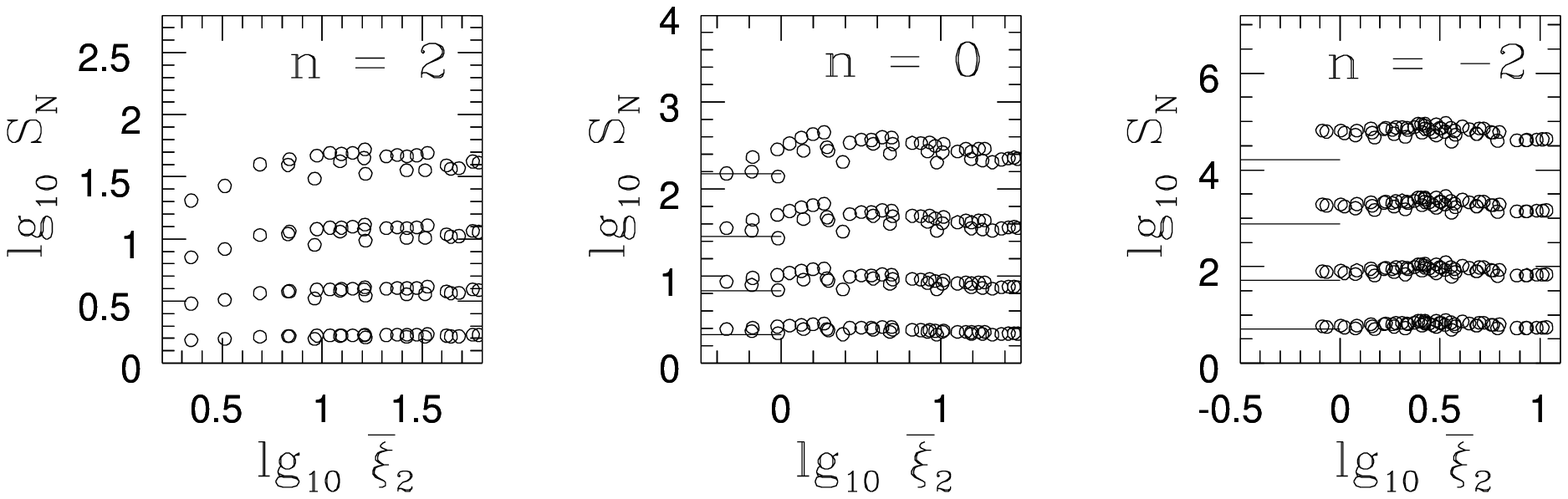}
}
\caption{The computed $S_p$ parameters are plotted against $\bar{\xi}_2$
for $n=2$, $n=0$ and $n=-2$ spectra in 2D. Solid lines represent 
predictions from perturbation theory. }
\end{figure*}

Once CPDF has been calculated one can use this information to
compute moments of this distribution and hence $S_p$ parameters
(eq.[4]). As recalled in the introduction 
these parameters have been studied extensively in recent past,
both analytically and computationally. In the limit 
$\sigma^2 \rightarrow 0$ and for Gaussian 
conditions they  are constant (Bernardeau 1992) 
and can be calculated using perturbation theory.
The resulting values of $S_3$ and $S_4$ are given in the
first section. Numerical results show good agreement with
theoretical predictions.

Of course these formula are not expected to be valid
in the intermediate and in the nonlinear regime.
Note however, that 
the spherical model might be of help to get insights into
those two regimes as pointed out by Mo \& White (1996),
Munshi \& Padmanabhan (1996).

In this paper we have developed a new method based on factorial moments (see 
Appendix) to compute $S_p$ parameters, with corrections due to finite volume
effects taken into account. Using this method we have computed $S_p$
parameters up to
$p = 6$ for both 2D and 3D and compared them with earlier results for
power law spectra. We found that finite volume corrections
 are extremely important as scale of nonlinearity increases. They are 
most important for spectra with lot of large scale power e.g. $ n = -2 $.
 Comparison of our results with earlier 
results of Colombi et al. (1996) where different method of volume
corrections were applied  shows reasonable agreement within 
the dynamical range studied by us. We however find that the $S_p$
coefficients in the non-linear regime are reasonably constant: the
drift $S_p\propto \xib^{0.045\,(p-2)}$ seen by Colombi et al. (1996)
is no longer present. The difference in the finite volume corrections
explain why the agreement is better for
spectra with less large scale power. To do these corrections we are a
using a systematic method - see Appendix - which makes sure that all
normalization constraints are well respected. Note however that the 
corrections depend mainly on proper parameterization of $h(x)$. 
With increasing large scale power accurate determination of $h(x)$ and
$\sigma(N_c)$ becomes difficult. This may also partially 
explain the lack of agreement between our result and Colombi et al. (1996)
for $n = -2$ spectra.

The $S_p$ parameters were also studied by Lucchin et al. (1992) using
central moments. Our results shows good agreement for all spectra
with their results. Although finite volume corrections were not 
taken into account they used several realizations of the same spectra
 which made their results agree with ours where volume corrections were
 taken care of and only one realization has been used. Since variation
of $S_p$ was not studied for large values of $\bar {\xi}_2$ in   Lucchin et al. (1994)  it is difficult to compare their results with our results.

\subsection{The Void Probability Function}

\begin{figure*}
\protect\centerline{ 
\epsfysize = 2. truein
\epsfbox[29 527 588 711]
{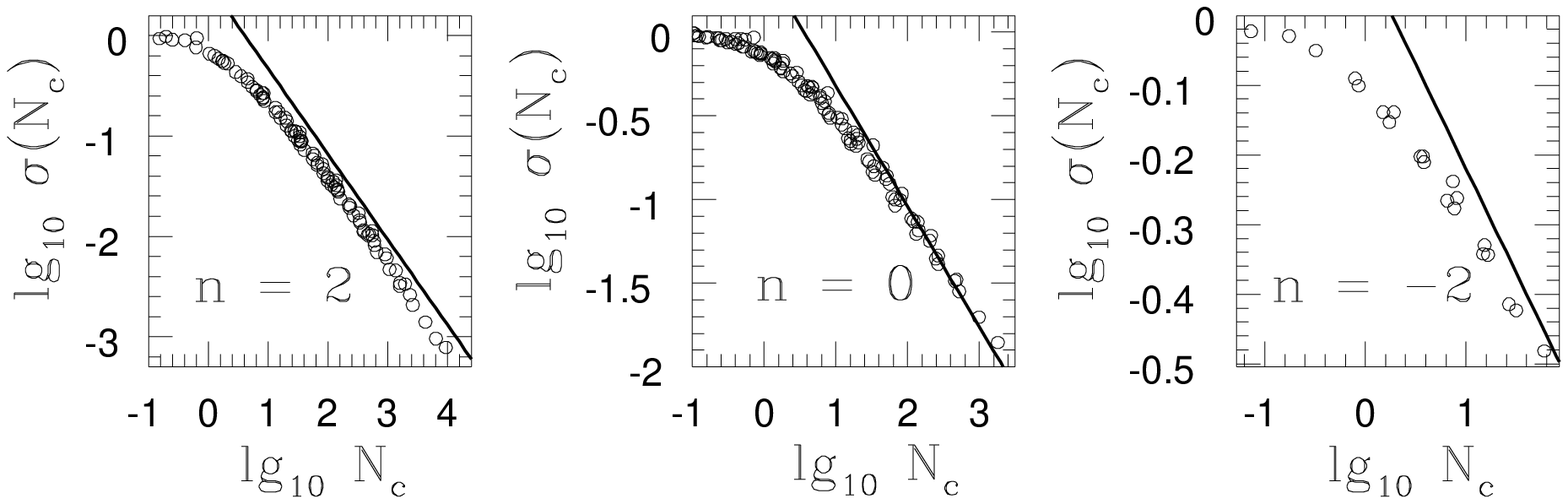}
}
\caption{The measured $\sigma(N_c)$ in 2D N-body simulations 
for $n=2$, $n=0$ and $n=-2$. The open symbols correspond to cases
where the variance is below unity, filled symbols for a variance above
unity. The slope of the solid lines are given in table 7.}
\end{figure*}

\begin{figure*}
\protect\centerline{ 
\epsfysize = 2.2 truein
\epsfbox[29 527 588 711]
{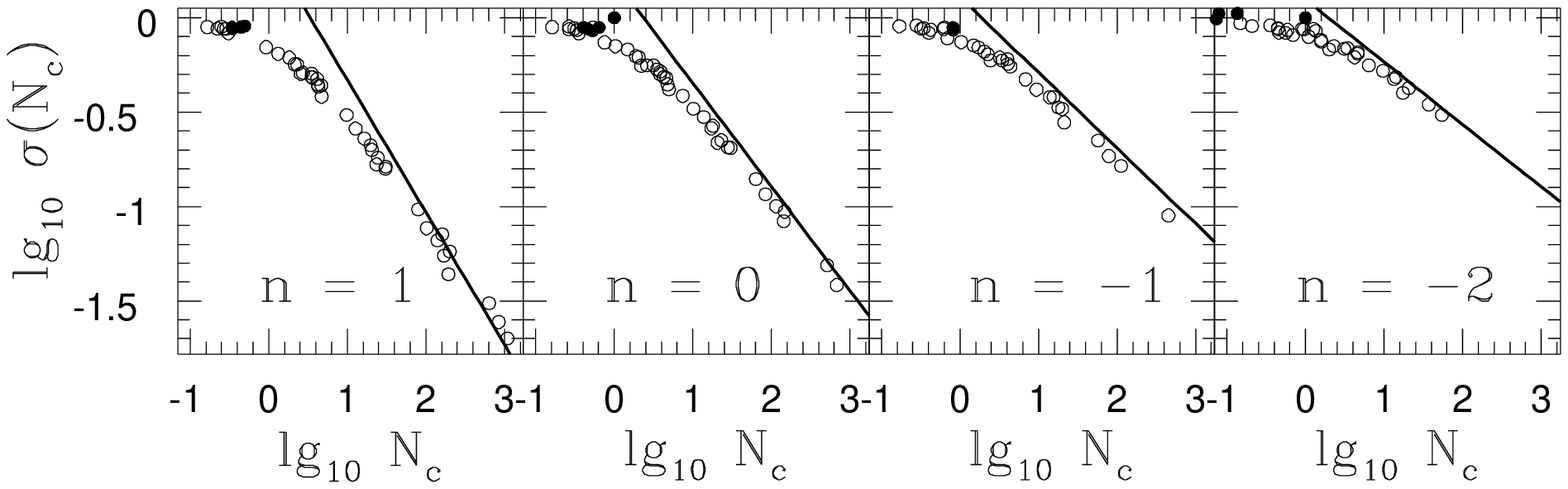}
}
\caption{The measured $\sigma(N_c)$ in 3D N-body simulations 
for $n=1$, $n=0$, $n=-1$ and $n=-2$. The open symbols correspond to cases
where the variance is below unity, filled symbols for a variance above
unity. The slope of the solid lines are given in table 8.}
\end{figure*}

In the previous section we have investigated mainly the large
density tails. 
As mentioned earlier the 
void probability function is directly related to the 
clustering properties but contains complementary information
in numerical measurements.
The scaling in the CPDF statistics is related to the 
scaling in $\sigma\equiv-\ln [P(0)]/(nV)$. Note that in the
absence of clustering, for a Poisson distribution, $\sigma$
is exactly unity and when some clustering is present the VPF
is expected to grow (it is more probable to find a void)
and hence $\sigma$ decreases. Moreover from the previous sections  
we know that  the function $\sigma$ 
when expressed as a function of $N_c$ is expected to have a 
power-law behavior for large $N_c$. For small $N_c$  it tends to the
Poisson limit, $\sigma(N_c)=1$.

 We have considered seven different epochs starting from the epoch when the 
grid scale goes nonlinear to the epoch when the box scale goes nonlinear. 
We have also considered different level of dilution to the full set of 
$512^2$ and $128^3$ data.

Different spurious effects affect the VPF differently, unlike CPDF
it is less affected by finite volume of the sample but it is very much 
affected by grid effects. Over-dense regions lose memory of the initial 
grid just 
after the first shell crossing but in under-dense regions the grid structure is
present till very late stages and these under-dense regions 
contribute more to VPF. It was shown that this effect is negligible 
 if we restrict ourselves to those cells where the conditions
 $P(0)>1/e$ is satisfied (Colombi et al. 1995). 
Grid effects are more significant for spectra with
more power on larger scales where as for spectra with lots of power in
smaller scales collapse of smaller objects erases the memory of initial 
grid very fast.

Scatter in plots increases from $n=2$ to $n=0$ and for 
$n=-2$ the $\sigma(N_c)$ diagrams reveal an significant
evolution with time. For $n = -2$ 
spectra larger and larger modes have more power and results are affected 
badly by finite volume corrections. Since we superpose all scales from 
quasi-linear to highly nonlinear regime in the same plots, any scatter
gives a measure of variation of $S_p$ parameters in these two regimes.
In earlier studies, it was noticed that variation of $S_p$ parameter 
from quasi-linear
to highly nonlinear regime increased with presence of large scale power.
This may explain the small scatter in our plot for $n$= 2 spectra. 
We do not see such a trend in our 3D analysis which may be due to the 
smaller available dynamic range. Computation of $\sigma(N_c)$ for very small
values of $\bar \xi_2$ is constrained by the restriction $P(0)>1/e$ which is 
to be satisfied for avoiding grid effect and discreteness effect.

\begin{table}
\caption{Measured values of $\omega$ for the 2D dynamics, 
Fig. 6, compared to the PT prediction, eq. (39).}
\label{tabsig2D}
\begin{tabular}{@{}lcccc}
                  &$n=2$&$n=0$&$n=-2$\\
$\omega^{\rm meas.}$&0.85&0.72&0.3\\
$\omega^{\rm PT}$&-&0.65&0.39
\end{tabular}
\end{table}

\begin{table}
\caption{Measured values of $\omega$ for the 3D dynamics, 
Fig. 7, compared to the PT prediction, eq. (38).}
\label{tabsig3D}
\begin{tabular}{@{}lcccc}
                  &$n=1$&$n=-1$&$n=0$&$n=1$\\
$\omega^{\rm meas.}$&0.7&0.55&0.4&0.33\\
$\omega^{\rm PT}$&1&0.75&0.6&0.5
\end{tabular}
\end{table}

The values of $\omega$ we get are given in table (\ref{tabsig2D}) 
for the 2D case and in table (\ref{tabsig3D}) 
for the 3D case where the errors on the measured values are about $0.05$.
It is interesting to see that PT predicts the
right trend for the $n$ dependence, although
there are significant discrepancies between the measured 
$\omega$ and the ones predicted by PT.

\section{Conclusion}

We have shown that PDF constructed from PT formalism
works extremely well for $ \sigma^2 \le 1 $.  Since all the loop corrections 
to $S_p$ 
parameters are neglected in this kind of approach it seems that loop 
corrections may not be very important in the perturbative regime for 
construction of PDF.

The constructed PDF is more accurate for spectra with less power on smaller 
scales. The existence of lots of small scale power produces a flow of power 
from smaller scale to larger scales which contradicts the basic assumptions 
of perturbation theory where density evolution at sufficiently large scales 
can always be described by linear theory.

For spectra with more power in larger scales the evolution is quite rapid and 
$P_N$ take the characteristic nonlinear power-law form quite early even when
$\sigma<1 $. 

We developed a method based on factorial moments to calculate higher
order correlation functions and used them to study evolution of $S_p$
parameters for power law spectra in 2D and 3D. Comparison with earlier
studies shows reasonable agreement.

\section*{Acknowledgment}
It is pleasure for D.M. to acknowledge Varun Sahni, his thesis supervisor
for constant encouragement and active support during the course of work.
 D.M. was financially supported by the Council of
Scientific and Industrial Research, India, under its SRF scheme.
D.M. also acknowledges financial support from CEA (Saclay) during his stay
there where most of the work was completed.
A.L.M. wishes to acknowledge the National Center for Super-computing 
Applications for support to perform the ensemble of simulations, and 
financial support  under NASA grant NAGW-2832.

\section*{Appendix}

\subsection{ P(N) and its continuous analogue $\Pi (\nu)$}
Using the relation between count in cell and void probability function
one can write,
\begin{equation}
P(N) = {(-1)^N \over N!} {d \over d\mu^N} \exp \bigg( - {\phi(\mu N_c) 
\over \xi_2} \bigg)\bigg |_{\mu = 1},
\end{equation}
which can also be written in the following form
\begin{equation}
P(N) = {1 \over 2\pi i}  \int { d \lambda \over \lambda^{N+1}}
\exp \bigg( - { \phi ((1 - \lambda) N_c) \over \xi_2} \bigg).
\end{equation}
The above integral has to be evaluated along a contour around $\lambda = 0$.

We then  define  the function $\Psi(t) = - \phi(-t)$  with which
the function $\Pi(\nu)$ is defined,
\begin{equation}
\exp \big ({ \Psi(t) \over \bar \xi_2 }\big ) = 
\int_0^{\infty} d \nu \exp( {\nu t \over N_c} ) \Pi( \nu).
\end{equation}
Using the definition of $P(N)$ now one can easily show that 
\begin{equation}
P(N) = \int_0^{\infty} d \nu { \exp(-\nu) \nu^N \over N! } \Pi( \nu).
\end{equation}

Therefore $\Pi(\nu)$ can be viewed as the continuous limit of 
$P(N)$ in the limit of large number densities. This can be seen 
by change of variable $\lambda = 1 + t/N_c$ in equation (2)
and then taking the limit $N_c \rightarrow \infty$ and 
$N \rightarrow \infty$ with the ratio  
$N / N_c$ remaining finite which gives $ P(N) = \Pi(N)$.
More precisely $P(N)$ is the convolution of the function
$\Pi(\nu)$ with a Poisson distribution describing
the shot noise effects. 

\subsection{ Factorial moments and $S_p$ parameters}
The factorial moments of $P(N)$ can be related with moments of $\Pi(\nu)$ 
by the following expression.
\begin{equation}
\sum_{N=0}^{\infty} N(N-1)... (N-p+1)P(N)= \int_0^{\infty} \nu^p
\Pi(\nu) d\nu.
\end{equation}

Now let us expand $\Psi(t)$ in a power series of $t$, 
$\Psi(t) = \sum_1^{\infty} \Psi_p t^p$.
One can convince himself that the coefficients $\Psi_p$ have the following 
relation with the $S_p$ parameters; $S_p = p! \Psi_p$.
Since in realistic scenarios, $S_p$ behaves as $p!$ the $\Psi_p$ 
coefficient are expected to be of order unity.

Similarly one can define the normalized factorial moments of $P(N)$ by the 
following expression
\begin{equation}
\Sigma_p = {\bar \xi_2 \over p! N_c^p} \sum N(N-1)... (N-p+1)P(N).
\end{equation}
These numbers are also of order unity. One can define a generating function 
$\Sigma(t)$ for $\Sigma_p$ parameters $\Sigma(t) = \sum_1^{\infty} \Sigma_p t^p$. We have the following expression connecting these two generating functions
$\Psi(t)$ and $\Sigma(t)$,

\begin{equation}
1 + {\Sigma(t) \over \bar \xi_2} = \exp ( {\Psi(t) \over \bar \xi_2} ),
\end{equation}
which can also be written as,
\begin{equation}
\Psi(t) = \bar \xi_2 \ln ( 1 + {\Sigma(t) \over \bar \xi_2} ).
\end{equation}
Expanding the above relation in powers of $1/\bar \xi_2$ we can write,
\begin{equation}
\Psi(t) = \Sigma(t) - {1 \over 2} { \Sigma^2(t) 
\over \bar \xi_2} + { 1 \over 3}{ \Sigma^3(t) \over
\bar \xi_2^3 } ..
\end{equation}

One can then write down the expressions of $\Psi_p$'s as a function of 
$\Sigma_p$'s. We present some lower order relations here,
\begin{equation}
\Psi_1 = \Sigma_1 
\end{equation} 
\begin{equation}
\Psi_2 = \Sigma_2 - {1 \over 2} { \Sigma_1^2 \over \bar \xi_2}
\end{equation}
\begin{equation}
\Psi_3 = \Sigma_3 - { \Sigma_1 \Sigma_2 \over \bar \xi_2} + {1 \over 3}
{ \Sigma_1^3 \over \bar \xi_2^2 }
\end{equation}
\begin{equation}
\Psi_4 = \Sigma_4 - { (\Sigma_1 \Sigma_3 + {1 \over 2} \Sigma_2^2)} {1  \over \bar \xi_2} + 
{ \Sigma_1^2 \Sigma_2 \over \bar \xi_2^2 } - {1 \over 4}
{ \Sigma_1^4 \over \bar \xi_2^3} 
\end{equation}
\begin{eqnarray}
\Psi_5 = &&\Sigma_5 - { (\Sigma_1 \Sigma_4 +  \Sigma_2 \Sigma_3)} {1  \over \bar \xi_2}  \nonumber \\
&&{+~(\Sigma_1^2 \Sigma_3 + \Sigma_1 \Sigma_2^2)} {1  \over \bar \xi_2^2} - { \Sigma_1^3 \Sigma_2 \over \bar \xi_2^3 } + { 1 \over 5} { \Sigma_1^5 \over
\bar \xi_2^4}
\end{eqnarray}
\begin{eqnarray}
\Psi_6 = && \Sigma_6 - { (\Sigma_1 \Sigma_5 + \Sigma_2 \Sigma_4 {1 \over 2} \Sigma_3^2)} {1  \over \bar \xi_2} \nonumber \\
&& +~{ (\Sigma_1^2 \Sigma_4 + 2 \Sigma_1 \Sigma_2 \Sigma_3 +  {1 \over 3} \Sigma_2^3)} {1  \over \bar \xi_2^2} \nonumber \\
&&-~{ (\Sigma_1^3 \Sigma_3 +  {3 \over 2} \Sigma_1^2\Sigma_2^2)} {1  \over \bar \xi_2^3} + { \Sigma_1^4 \Sigma_2 \over \bar \xi_2^4 } - { 1\over 6}{ \Sigma_1^6 \over \bar \xi_2^5 } 
\end{eqnarray}

\noindent
These formulae are valid for arbitrary values of $\bar \xi_2$.
Note that of course in the limit $\bar\xi\to\infty$ 
we have $\Sigma_p=S_p$.

\subsection{ Minimum and maximum length scales to be probed} 

The particle positions in simulations sample an underlying continuous field. 
To extract this field from actual output, one has to work with sufficiently 
large cells, so that the 
field points appear continuous.  The condition is given by $N_c \gg 1 $.
In practice
if $l_c$ is the scale where $N_c=1$, scales of few times $l_c$ start to be 
usable and obviously the code resolution provides the minimum length scale 
that can be probed. Choice of maximum length scale is slightly arbitrary 
as larger and larger cells starts finite volume effects.
These effects can be visualize however with irregularities in the
shape of the measured PDF-s.

\subsection{Number of trials}

 The typical distances between particles in a cluster is $l_c$. If the 
sample is divided in cells of size $l$ some information at scales smaller 
than $l$ is erased. To recover all informations available in the sample 
one has to use many grids displaced by a distance $l_c$ from each other. 
As mentioned earlier in 3D there are $(l/l_c)^3$ grids and $(L/l_c)^3$ 
cells, but only $(L/l)^3$ cells are completely independent. 

Events corresponding to $P(N)$'s smaller than $(l/L)^3$ will be either 
over represented in case there is one such event in the sample, or under 
represented in case there is none in the sample. So it is clear that one 
has limited access to
$P(N)$'s smaller than the above value. From the exponential decay of 
$\Pi(\nu) \propto \exp( - \nu/ \nu_*)$ with $\nu_* = x_*N_c$ one gets 
a limit for $N$ above which the information about $P(N)$ is not 
contained in the sample.

\begin{equation}
N_{max} \approx x_* N_c \ln(L/l_c)^3
\end{equation}
A more realistic parameterization of $\Pi(\nu)$ for $\nu \ll N_c$ 
would be

\begin{equation}
\Pi(\nu) \propto {1 \over N_c \bar \xi_2 } h( {\nu \over N_c})
\end{equation}
which leads to the formula 

\begin{equation}
N_{max} \approx x_*N_c \ln \bigg[\bigg( {L \over l} \bigg)^3 
{ \sqrt \pi x_* \over
\nu_*^2 \bar N } \ln^{5/2} \bigg( {L \over l } \bigg ) \bigg] 
\end{equation}

For $N>N_{max}$, $P(N)$ abruptly drops to zero due to the finiteness
of the sample. The moments $\Sigma_p$ are dominated by values of $\nu$ 
that can be inferred  to be

\begin{equation}
\nu_{ \Sigma_p}  \approx ( p - 5/2) \nu_*;~~~~~~~~ p\ll5/2
\end{equation}

These coefficients are not known for $\nu_{\Sigma}>N_{max}$ that is 
for the simple for (19) $p$ must satisfy

\begin{equation}
p < 5/2 + 3 \ln(L/l_c)
\end{equation}

$P(N)$ whose value is between $(l/L)^3$ and $(l_c/L)^3$ contains systematic
wiggles due to the fact that only $(L/l)^3$ cells out of $(L/L_c)^3$ cells are fully independent but averages of $P(N)$'s such as the moments (5) are 
less sensitive to this.  It is nevertheless better to use this information
than to drop it. With $(L/l)^3$ trials $P(N)$ will become inaccurate much
earlier. $\Sigma_p$ are systematically under estimated for large $p$ due to
abrupt drop in $P(N)$ for $N>N_{max}$.

\subsection{ Correction for finite volume effect}

A simple way to correct $\Sigma_p$ is to use the form (17) to supplement the missing information at large $N$, and to use $P(N)$ for $N < N_{max}$.

\begin{eqnarray}
P^{corr}(N)&=&\big( 1 - {a \over \bar \xi_2} - b {( N - N_c ) \over N_c} \big )
P^{com} (N);~~ N<N_{max} \nonumber \\
P^{corr}(N) &\approx& \Pi( \nu);~~ N>N_{max} 
\end{eqnarray}

It is clear that the corrected P(N) so constructed have to be normalized 
properly before using it to calculate $S_p$ parameters.

Using the constrain $\sum P(N) = 1 $ we  get $a = H_0$ and $\sum NP(N) = 1 $
gives us $b = H_1$. 
Where we have used the following notation 

\begin{equation}
H_p =  \int_{N_{max}/N_c}^{\infty} x^p h(x) dx 
\end{equation}

Where we have neglected the corrections of order $1/ \bar \xi_2$.
The corrected $N_c$ now can be written as

\begin{equation}
N_c^{corr} = ( 1 - 6 H_1 \Sigma_3 + H_2 ) N_c^{comp}
\end{equation}
and finally the corrected $\Sigma_p$'s are of the form

\begin{equation}
\Sigma_p^{corr} = {{ \Sigma_p^{comp} - (p+1)H_1 \Sigma_{p+1}^{comp} + 
{1 \over p! } H_p } \over { (1 - 6 H_1 \Sigma_3 + H_2 )^{p-1} }}
\end{equation}

 Corrected $S_p$ parameters can now
be recovered by using relation between $\Psi_p$ and $\Sigma_p$ 
as described already. This way to correct has the advantage of being 
rather simple and preserve all normalizing conditions.

This method of calculating $S_p$ parameter is an alternative to the method
generally used based on central moments. 
\end{document}